\documentclass[aps,reprint,amsmath,amssymb,superscriptaddress,longbibliography,nofootinbib,showkeys]{revtex4-1}
\pdfoutput=1
\usepackage{graphicx,xcolor}
\usepackage{slashed}
\usepackage{float}
\usepackage[
    colorlinks,
    linkcolor=green!30!blue,           
    citecolor=red,            
    filecolor=green!30!blue,        
    urlcolor=green!30!blue,         
    hyperfootnotes]{hyperref}
\usepackage{bm}
\usepackage{url}
\usepackage{accents}
\usepackage[paperwidth=215.9mm,paperheight=279.4mm,centering,hmargin=2.1cm,vmargin=2cm]{geometry}

\usepackage{tcolorbox}
\usepackage{multirow}
\usepackage{tabularx}
\usepackage{array}
\usepackage{colortbl}
\tcbuselibrary{skins}
\newcolumntype{Y}{>{\raggedleft\arraybackslash}X}
\tcbset{tab1/.style={fonttitle=\bfseries\large,fontupper=\normalsize\sffamily,
colback=white,colframe=black,colbacktitle=Salmon!40!white,
coltitle=black,center title,freelance,frame code={
\foreach \n in {north east,north west,south east,south west}
{\path [fill=red!75!black] (interior.\n) circle (3mm); };},}}

\tcbset{tab2/.style={enhanced,fonttitle=\bfseries,fontupper=\normalsize\sffamily,
colback=white,colframe=black,coltitle=black,center title}}

\begin{document}
\title{{The top threshold effect in the $\gamma\gamma$ production at the LHC}}
\author{Shashikant R. Dugad}
\email{shashi@tifr.res.in}
\affiliation{Department of High Energy Physics, Tata Institute of Fundamental Research,
1 Homi Bhabha Road, Mumbai 400 005, India}

\author{Pankaj Jain}
\email{pkjain@iitk.ac.in}
\affiliation{Department of Physics, Indian institute of Technology Kanpur, Kanpur 208 016, India}

\author{Subhadip Mitra}
\email{subhadip.mitra@iiit.ac.in}
\affiliation{Center for Computational Natural Sciences and Bioinformatics,
International Institute of Information Technology, Hyderabad 500 032, India}

\author{Prasenjit Sanyal}
\email{psanyal@iitk.ac.in}
\affiliation{Department of Physics, Indian institute of Technology Kanpur, Kanpur 208 016, India}

\author{Ravindra K. Verma}
\email{ravindkv@iitk.ac.in, ravindra.verma@tifr.res.in}
\affiliation{Department of Physics, Indian institute of Technology Kanpur, Kanpur 208 016, India}
\affiliation{Department of High Energy Physics, Tata Institute of Fundamental Research,
1 Homi Bhabha Road, Mumbai 400 005, India}


\begin{abstract}
We compute the top quark threshold contributions 
to the $\gamma\gamma$ production at the LHC. 
They appear when the invariant mass of the photon pair, $M_{\gamma\gamma}$ 
just exceeds two times the mass of the top quark and induce 
some feature in the $M_{\gamma\gamma}$ distribution. We determine the magnitude of this threshold effect
and characterize this feature with a simple empirical fitting function to show that it is possible to observe this effect at the LHC in future.  We also explore some possible improvements
that may enhance its significance.
\end{abstract}
\date{\today}
\maketitle
\section{Introduction}
\label{sec:intro}
\noindent
The $\gamma\gamma$ pair production in proton-proton colliders such as the 
LHC plays a very important role in the search for new physics. 
Recently this channel {had} attracted considerable attention due to
a potential hint of new physics. The  2015 LHC data showed an excess over the Standard Model (SM) expectations around
$M_{\gamma\gamma}=750$ GeV where $M_{\gamma\gamma}$ is  the invariant mass of the photon pair \cite{ATLAS:2015,ATLAS:2016,CMS:2015dxe}.
However,  the peak is absent in the latest record with improved statistics \cite{ATLAS:2016eeo,CMS:2016crm} indicating that it was actually due to a statistical fluctuation.  Irrespective of the origin of the peak, the SM background to this channel should be computed as precisely as possible to capture the essential expected features in the background simulation.

The background generally shows a smooth behavior with respect to the invariant mass of the photon pair ($M_{\gamma\gamma}$), as indicated, for example,
by the background-only fits obtained by the ATLAS collaboration \cite{ATLAS:2015,ATLAS:2016eeo}. 
However, at the threshold of production of a particle, this smooth feature would get modulated. 
In this paper, we primarily look at one such effect originating within the SM, namely, the top quark threshold effect that appears around $M_{\gamma\gamma}\approx$ 350 GeV.

The $\gamma\gamma$ pair production ($pp\to\gamma\gamma$) at the LHC gets contribution from
the quark-antiquark annihilation $(q\bar q\rightarrow \gamma
\gamma)$ at the leading order (LO) (with $\mathcal O(\alpha^2_{\rm ew})$ contribution to the cross-section). 
At the next-to-leading order (NLO) in QCD we get the $\mathcal O(\alpha^2_{\rm ew}\alpha_{s})$ contributions to the cross-section from virtual diagrams of quark-antiquark annihilation and real diagrams like quark-antiquark annihilation $(q \overline{q}\rightarrow g\gamma
\gamma)$ and  quark-gluon scattering $(q g\rightarrow q\gamma
\gamma)$ \cite{Binoth:1999qq}. At the next-to-next-to-leading order (NNLO) in QCD we have the $\mathcal{O}(\alpha^2_{ew}\alpha^2_{s})$ contribution to the cross-section from double-virtual, real-virtual and real-real diagrams \cite{Campbell:2016yrh}.

\begin{figure}[t]
\begin{center}
\scalebox{0.4}{\includegraphics*[angle=0,width=\textwidth,clip]{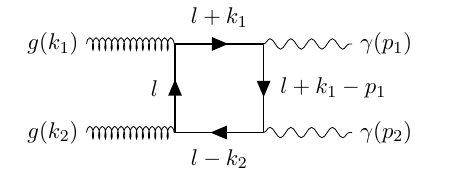}}\end{center}
\caption{A representative top-loop diagram contributing to $gg\to\gamma\gamma$. }
\label{fig:toploop}
\end{figure}

At the same order, {\it i.e.} $\mathcal O(\alpha^2_{\rm ew}\alpha^2_{s})$, another process opens up, namely, the fermion loop mediated gluon fusion process, $gg\to\gamma\gamma$ \cite{Bern:2001df,Bern:2002jx,Li:2011ye,Campbell:2016yrh,Chway:2015lzg}. It involves loop diagrams such as the box and the cross-box diagrams like the one shown in 
Fig.~\ref{fig:toploop} (box diagram). Though loop mediated, this process has no tree level part and hence we get a finite contribution from these loops. ( The two-loop matrix elements for the gluon fusion is given in \cite{Bern:2001df}.)
It is in this process where the top threshold effect appears from the destructive interference between the top loop diagrams containing on-shell top quarks and other diagrams containing light quark loops.\footnote{To avoid any confusion, throughout this paper we shall refer to all $pp\to \gamma\gamma$ processes except the box- and cross-box-diagram mediated $gg\to\gamma\gamma$ process (and its higher order corrections) collectively as $qX\to\gamma \gamma$ process computed up to different orders of QCD coupling.} Once $M_{\gamma\gamma}$
exceeds two times the mass
of the top quark, $m_t\approx 173$ GeV, the gluon fusion process gets contribution 
from on-shell tops in the box loop, creating a dip in the invariant mass distribution 
\cite{Li:2011ye,Campbell:2016yrh,Chway:2015lzg}.

The precise nature of the threshold effect can be seen explicitly in Fig.~4 of \cite{Li:2011ye}. It 
shows the ratio of the cross-sections of the gluon
fusion process computed with $m_t=173$ GeV to that obtained
by setting $m_t = \infty$ at the 14 TeV LHC. In other words, it shows the correction due to the top loop to the gluon fusion process. 
This ratio is found to be equal to unity for $M_{\gamma\gamma}<200$ GeV.
As $M_{\gamma\gamma}$ increases, the ratio starts to decrease and
shows a sudden dip at about $M_{\gamma\gamma}=2m_t$. After the dip, it rises 
smoothly and eventually
saturates to a value of approximately 1.75 for $M_{\gamma\gamma}\approx 1600$
GeV. Note, however, there is nothing special about the top quark or about this process, similar dips are also predicted at the threshold of each new particle in the light by light scattering \cite{Bohm:1994sf}. 

Though this threshold effect exists, a priori it is not  clear whether 
it can be observed since the gluon-gluon
fusion gives a sub-dominant contribution to the two photon production
\cite{Aurenche:1985yk,Bailey:1992br,PhysRevD.47.2735,PhysRevD.49.1486,Binoth:1999qq,Anastasiou:2002zn,Catani:2011qz}. 
As mentioned, the leading order contributions arise from the quark anti-quark
annihilation process. 
Fortunately, the {higher order} contributions are
relatively large for the $\gamma\gamma$ pair production 
\cite{Catani:2011qz}. 
Mainly because of the large gluon parton density function (PDF), the gluon fusion process, although higher order in
strong coupling in comparison to the quark anti-quark annihilation
process,  is not negligible and gives a significant contribution to the cross-section
\cite{Bern:2001df,Bern:2002jx,Catani:2011qz}. 
This may be further enhanced by imposing some kinematic
cuts. An even higher order contribution
to this, classified as {N$^3$LO}, has also been computed 
\cite{Bern:2002jx}. It is found to be small but not negligible.

It is intriguing that the {2015} ATLAS data \cite{ATLAS:2015} {showed} a hint of a dip at 
$M_{\gamma\gamma}\approx 2 m_t$, exactly where it is expected from the top threshold effect. The dip is not so obvious in the latest data set \cite{ATLAS:2016eeo}, but it is not in contradiction with the dip. In this paper, we numerically simulate NLO events for the $qX\to\gamma\gamma$ and LO events for the loop mediated $gg\to\gamma\gamma$ process to investigate the possibility of observing the top quark threshold effect at the 13 TeV LHC. 
Through a statistical analysis we aim to establish that though difficult, it is possible to observe this effect at the LHC in future.

Note that while observation of this phenomenon is interesting by itself, it 
may also be useful to understand the relative magnitudes of different 
contributions and the process in general. Theoretically, these have significant uncertainties due to the unknown higher order contributions. If the effect can
be observed with sufficient accuracy, it could provide another measurement of the top quark mass \cite{Kawabata:2016aya}.

Going beyond the top-quark effect, in general, studying such threshold
effects can be useful for new physics also. They can tell us about heavy particles contributing in loop processes where the final state particles are observed at the LHC.
For example, since it arises from the interference terms, any heavy particle (carrying electric and color charges) that can run in the loop of the
$gg \to\gamma\gamma$ process would lead to such a threshold effect.
Hence, even in absence of any direct detection of heavy particles at the LHC, observation (or non-observation) of any such effect
in the $\gamma \gamma$ spectrum could let us infer about 
heavy particles carrying non-zero electromagnetic
charge.

\section{Computations and results}
\label{sec:toploop}
\noindent
We generate events for both LO $gg\to\gamma\gamma$ and NLO $qX\to\gamma\gamma$ processes at the 13 TeV LHC in the {\sc MadGraph}5\_{\sc aMC}{\sc @NLO} \cite{Alwall:2014hca} environment with 
NN23NLO parton density functions (PDFs) \cite{Ball:2012cx}. We use  {\sc Pythia}6 \cite{Sjostrand:2006za} for parton showers (PS). Finally we pass the events through {\sc Delphes} 3.3.1 \cite{deFavereau:2013fsa}, a detector simulator,  with the default ATLAS card to generate realistic distributions.

It is possible to estimate the cross-sections of these processes more precisely with the parton level Monte Carlo code MCFM \cite{Boughezal:2016wmq,Campbell:2016yrh}.\footnote{$pp\to\gamma\gamma$ at NNLO QCD was first computed with 2gNNLO in Ref.~\cite{Catani:2011qz}.} It can compute the $qX\to\gamma\gamma$ process cross-section at $\mathcal O(\alpha^2_{\rm ew}\alpha^2_{s})$ (NNLO) and the $\mathcal O(\alpha^2_{\rm ew}\alpha^3_{s})$ (NLO) correction to the $gg\to\gamma\gamma$ process. In our estimations, we include the effects of these higher order corrections to the production processes in the form of overall $K$-factors. We scale the NLO $qX\to\gamma\gamma$ cross-section (obtained from {\sc MadGraph}) by 
\begin{eqnarray}
K^{qX}_{\rm NNLO}=\frac{\sigma_{qX\to\gamma\gamma} \mbox{ obtained at }\mathcal O(\alpha^2_{\rm ew}\alpha^2_{s})}{\sigma_{qX\to\gamma\gamma} \mbox{ obtained at }\mathcal O(\alpha^2_{\rm ew}\alpha_{s})} = 1.70\label{eq:kpp}
\end{eqnarray}
and the LO $gg\to\gamma\gamma$ cross-section by,
\begin{eqnarray}
K^{gg}_{\rm NLO}=\frac{\sigma_{gg\to\gamma\gamma} \mbox{ obtained at }\mathcal O(\alpha^2_{\rm ew}\alpha^3_{s})}{\sigma_{gg\to\gamma\gamma} \mbox{ obtained at }\mathcal O(\alpha^2_{\rm ew}\alpha^2_{s})}=1.48\,.
\label{eq:kgg}
\end{eqnarray}
The numerators in Eqs.~\eqref{eq:kpp} and \eqref{eq:kgg} are computed with the NN23 NNLO PDF sets. These $K$-factors are estimated for the region of the phase-space where $E^\gamma_{\rm T}>40$ GeV and $M_{\gamma\gamma} \geq 200$ GeV.  For our calculations we have considered photons with $E^\gamma_{\rm T}>40$ GeV and $|\eta^\gamma| < 2.5$ only. The photons are isolated using a smooth cone isolation prescription \cite{Frixione:1998jh} with $\epsilon_\gamma=1$, $n=1$, $\delta_0=0.4$ (the choice of the isolation parameters are motivated by Ref.~\cite{Frixione:1998jh}).
We consider dynamic renormalization and factorization scales and set them as $\mu_{\rm R}=\mu_{\rm F}=M_{\gamma\gamma}$.

\subsection{Gluon Fusion ($gg\to\gamma\gamma$)}\label{eq:Sec1}
\begin{figure}[t]
\begin{center}
\scalebox{0.37}{\includegraphics*[angle=0,width=\textwidth,clip]{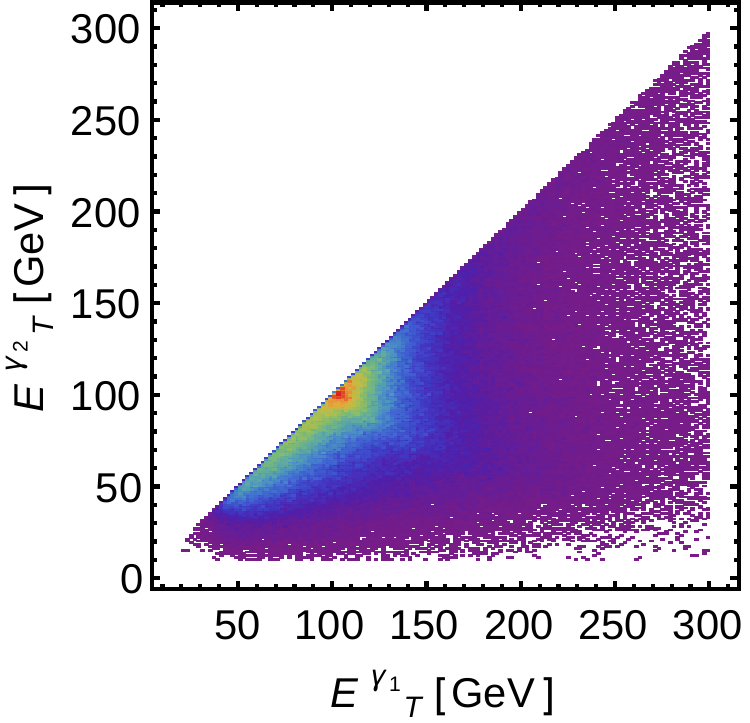}}
\scalebox{0.078}{\includegraphics*[angle=0,width=\textwidth,clip]{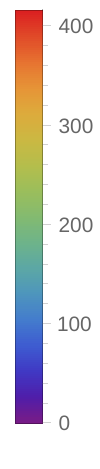}}
\end{center}
\caption{ Correlation among the transverse energy of the two photons coming from the gluon fusion process.The photon density per $10\times10$ GeV$^2$ bin is given by the color bar. 
}
\label{fig:2dcor}
\end{figure}

\noindent
For $E^\gamma_{\rm T}>40$ GeV and $M_{\gamma\gamma} \geq 200$ GeV, the LO cross-section for the gluon fusion is about $162$ fb which roughly increases to $240$ fb after multiplying with $K^{gg}_{\rm NLO}$ [Eq.~\eqref{eq:kgg}].  The ATLAS analysis~\citep{ATLAS:2015} imposes the following additional cuts on the photons,
\begin{eqnarray}
E^{\gamma_1}_{\rm T} > 0.4 M_{\gamma\gamma},\quad
E^{\gamma_2}_{\rm T} > 0.3 M_{\gamma\gamma}.\quad\label{eq:oATLAScuts}
\end{eqnarray}
Since we look at $M_{\gamma\gamma} \geq 200$ GeV only, applying the ATLAS cuts will ensure all events have $E^{\gamma_1}_{\rm T} > 80$ GeV and $E^{\gamma_2}_{\rm T} > 60$ GeV. Fig.~\ref{fig:2dcor} (where we have shown the correlation among the transverse momentum/energy of the two photons) indicates that it is possible to keep more gluon-fusion events by relaxing these cuts.%
\footnote{Apart from the events in the low $E^{\gamma}_{\rm T}$ region, the ATLAS cuts \eqref{eq:oATLAScuts} will also eliminate any event with $M_{\gamma\gamma} \sim 350$ GeV (the region of our interest), if it has $E^{\gamma_1}_{\rm T}$ between $105$ and $140$ GeV and $E^{\gamma_2}_{\rm T}$  greater than $105$ GeV (approximately) -- a region which is quite densely populated in Fig.~\ref{fig:2dcor}.}
Hence, we choose the following cuts:
\begin{eqnarray}
E^{\gamma_1}_{\rm T} \geq E^{\gamma_2}_{\rm T} \geq 0.25 M_{\gamma\gamma}.\quad
\label{eq:ATLAScuts}
\end{eqnarray}
We find that the ratio of $gg$ to $qX$ events does not change substantially by this change in cuts.
The cross-section in the gluon fusion channel reduces to about $130$ fb 
once these cuts are imposed.

\begin{figure}[t]
\begin{center}
\scalebox{0.45}{\includegraphics*[angle=0,width=\textwidth,clip]{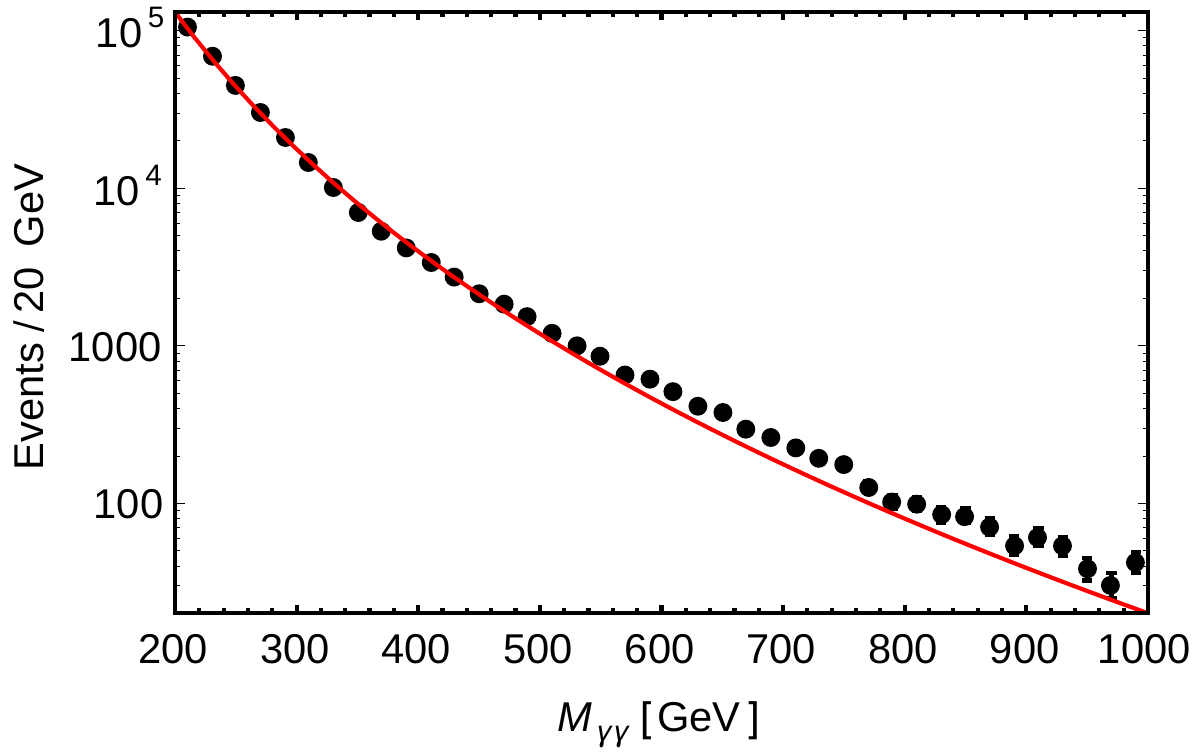}}
\scalebox{0.45}{\includegraphics*[angle=0,width=\textwidth,clip]{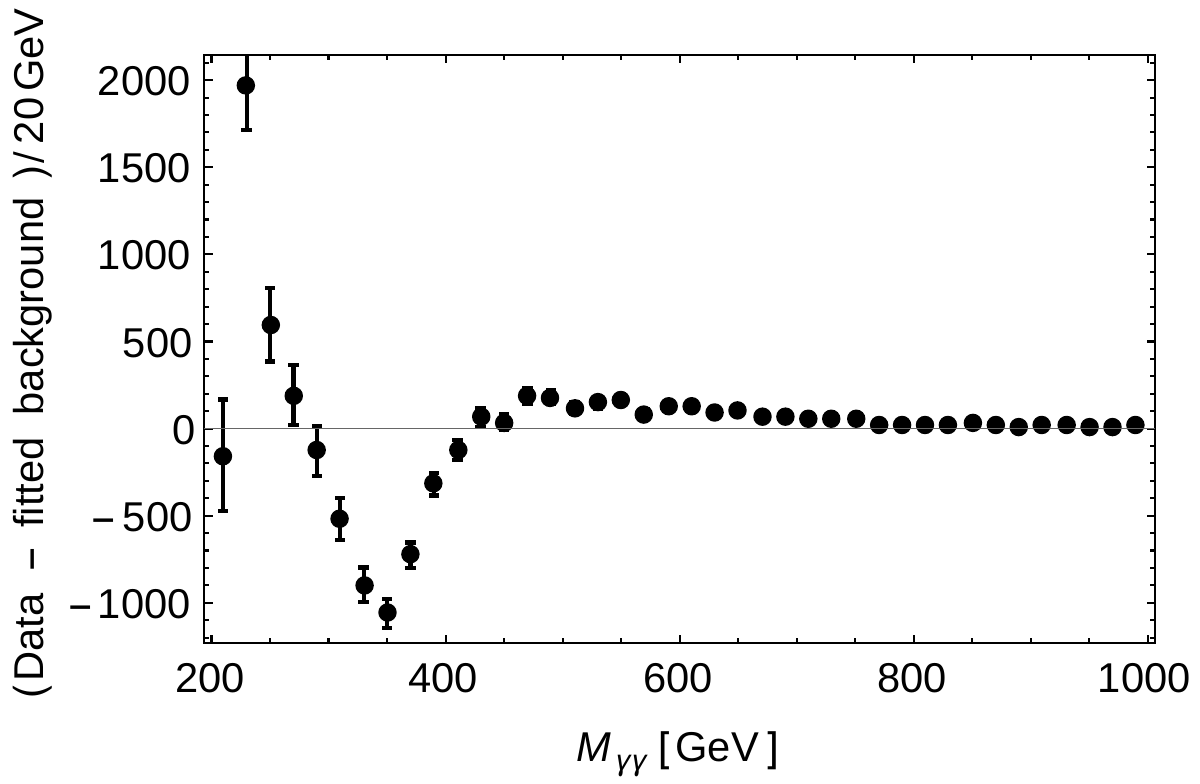}}
\scalebox{0.45}{\includegraphics*[angle=0,width=0.95\textwidth,clip]{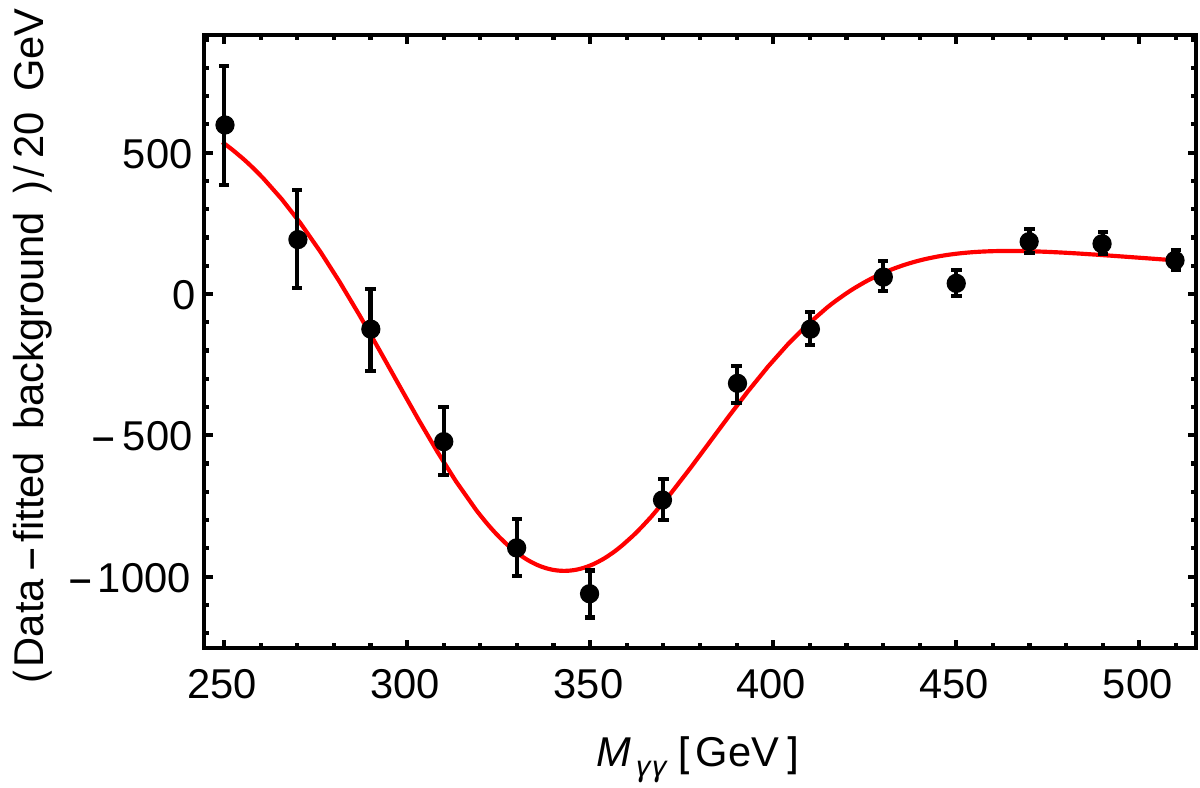}}
\end{center}
\caption{ The invariant mass distribution of the two photons from the gluon fusion process. These events are obtained after applying the cuts defined in Eq.~\eqref{eq:ATLAScuts}. The solid line in the top plot shows a smooth fit of
the binned events in the $200$ GeV $\leq M_{\gamma\gamma}\leq$ $1000$ GeV range with the smooth ATLAS fitting function [Eq.~\eqref{eq:fitform}]. The middle plot shows the difference
between the simulated events and the smooth fit.
A clear dip is seen at $M_{\gamma\gamma}\approx 2m_t$. In the range $250 \leq M_{\gamma\gamma} \leq 510$ GeV the difference is fitted with the function shown in Eq.~\eqref{eq:gg_fitform} that captures the dip better as shown in the bottom plot. 
}
\label{fig:thres}
\end{figure}

We show the invariant mass distribution obtained from the $gg\to\gamma\gamma$ process
in Fig.~\ref{fig:thres} (top plot). To show the threshold effect prominently in this plot, we have generated a large number of events ($600$K) with $M_{\gamma\gamma} \geq 200$ GeV in the gluon fusion channel alone and then applied the cuts defined in Eq.~\eqref{eq:ATLAScuts} on these events. 
This is useful in order to precisely determine the shape of this dip. In 
full analysis we will use only a smaller number of events which can 
be obtained in a reasonable time scale at the LHC. 
The red line in the top panel is a smooth fit to the simulated events in the $200$ GeV $<M_{\gamma\gamma}<$ $1000$ GeV range. 
The fitting function is taken to be
of the same form as used by ATLAS~\citep{ATLAS:2015},
\begin{equation}
f_0(x) = (1-x^{1/3})^bx^{a_0}, \quad x = M_{\gamma\gamma}/\sqrt{s}.
\label{eq:fitform}
\end{equation}
The top threshold effect is very clearly visible in the middle panel of Fig.~\ref{fig:thres} where we have shown the difference between the simulated events and the smooth fit. The errorbars shown represent the the square-root of the number of events in each bin ($\sqrt N$).
We see a clear dip in the cross-section approximately at the position of
twice the top quark mass.\footnote{Note that the fit is not very precise at the low values of $M_{\gamma\gamma}$ ($\sim 200$ GeV), but this happens because of the limitations of the simple form of the fitting function in Eq.~\eqref{eq:fitform}.} It originates from the destructive interference between the top loop diagrams containing on-shell top quarks and other diagrams containing light quark loops. We fit the dip with the following function,
\begin{eqnarray}
g_0(x) &=& A\left[- \exp\Big\lbrace-\frac{(x-x_0)^2}{2\sigma_{g}^2}\Big\rbrace\right. \nonumber \\
&&\quad\quad +\left.\mathcal R\Big\lbrace \exp\left(-\frac {x}{ \sigma} \right)-\left(\frac{\sigma}{x}\right)^4\Big\rbrace\right] 
\label{eq:gg_fitform}
\end{eqnarray}
in the range $250 \leq M_{\gamma\gamma} \leq 510$ GeV (see the bottom panel of Fig.~\ref{fig:thres}). The parameter $x$ is defined in Eq. \ref{eq:fitform}. 
Roughly, the Gaussian part accounts for the dip and the other term dictates the steep behavior at lower $M_{\gamma\gamma}$. The parameter $x_0$ indicates the location of the dip. The fit gives $x_0 = 2.6077\times 10^{-2}$ which corresponds to $M_{\gamma\gamma} = 339$ GeV which is roughly twice the top quark mass. The width of the dip is given by $\sigma_g = 3.3169 \times 10^{-3}$ which is equivalent to a width of $43.1$ GeV. The parameter $\sigma$ becomes $9.4362\times10^{-3}$. The $\chi^2$ per degree of freedom for this fit is about $1.2$. The relative factor $\mathcal R= 6.8338$. 
The overall normalization depends on the number of events generated and is specified below for the final fit.

\subsection{Other Processes ($qX\to\gamma\gamma$)}\label{eq:Sec2}

\noindent
As mentioned before, the total $qX\to\gamma\gamma$ cross-section is much larger than that of the $gg\to\gamma\gamma$ process. For good statistics we have generated $8$ million events with $E_{\rm T}^\gamma\geq 40$ GeV in the $qX\to\gamma\gamma$ channel. This corresponds to a luminosity of about $346$ fb$^{-1}$. After applying the cuts defined in Eq.~\eqref{eq:ATLAScuts}, the total $qX\to\gamma\gamma$ cross-section comes down to $1718$ fb which is still about thirteen times larger than the corresponding $gg\to\gamma\gamma$ cross-section. 

\begin{figure}[t]
\begin{center}
\scalebox{0.5}{\includegraphics*[angle=0,width=\textwidth,clip]{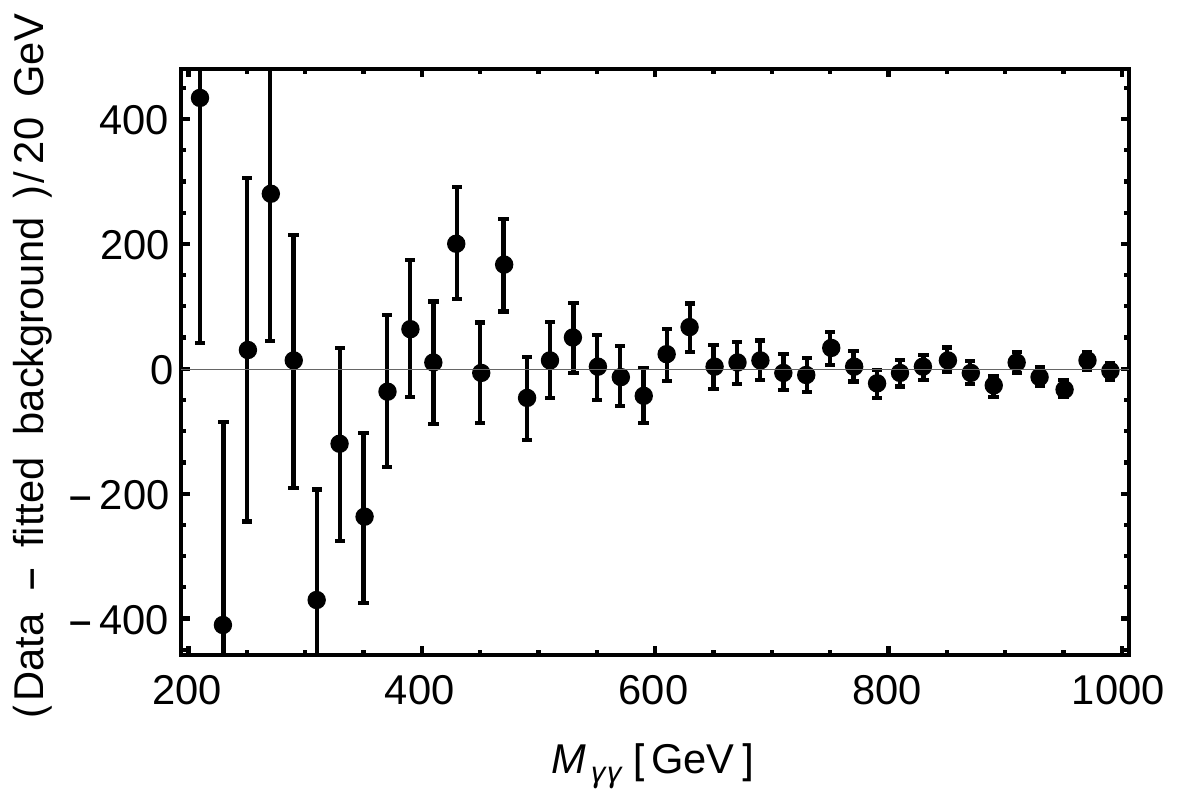}}
\end{center}
\caption{The difference between the binned events and their smooth fit by the ATLAS fitting function [Eq.~\eqref{eq:fitform}] obtained for the $qX\to\gamma\gamma$ processes. 
}
\label{fig:thres2}
\end{figure}

We fit these events with the ATLAS fitting function [Eq.~\eqref{eq:fitform}]. In Fig.~\ref{fig:thres2}  we show the difference between the fit and the events. Notice that, even with such high number events, the difference plot shows fluctuations indicating small deviations from the fit
in the region of our interest namely $250 \leq M_{\gamma\gamma} \leq 510$ GeV.

\subsection{Combined Processes ($gg\to\gamma\gamma + qX\to\gamma\gamma$)} \label{eq:Sec3}

\begin{figure}[t]
\begin{center}
\scalebox{0.5}{\includegraphics*[angle=0,width=\textwidth,clip]{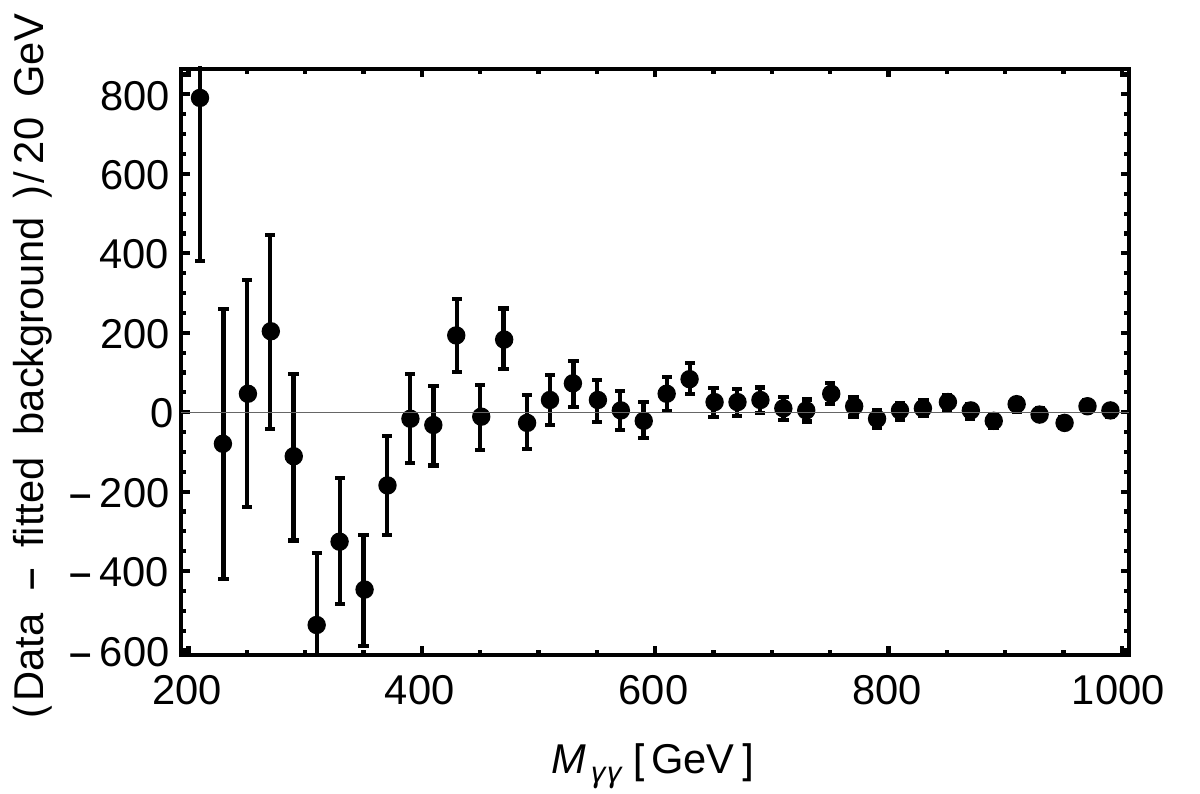}}
\scalebox{0.48}{\includegraphics*[angle=0,width=0.95\textwidth,clip]{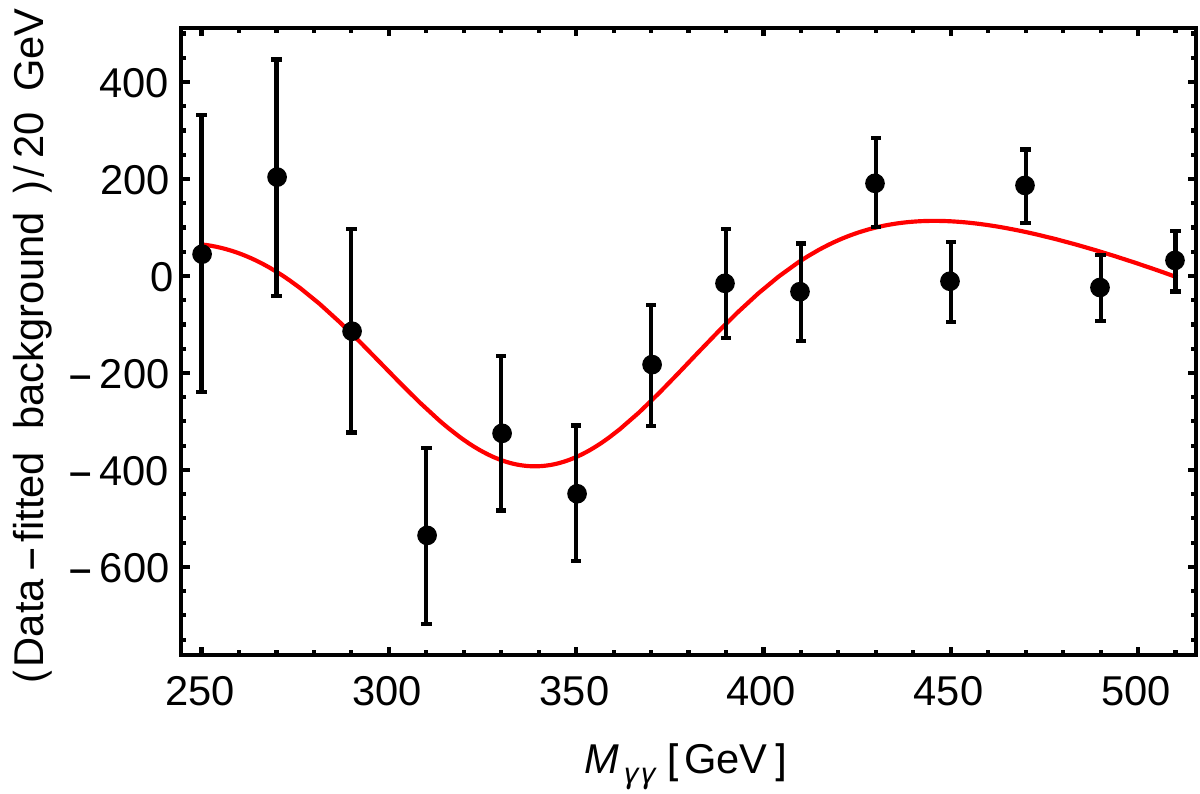}}
\end{center}
\caption{The top plot shows the difference between the binned events and their smooth fit by the ATLAS fitting function [Eq.~\eqref{eq:fitform}] obtained for the combined $gg\to\gamma\gamma + qX\to\gamma\gamma$ processes. 
The bottom plot shows the difference modeled with the function defined in Eq.~\eqref{eq:combined_fitform} in the range $250 \leq M_{\gamma\gamma} \leq 510$ GeV. 
}
\label{fig:thres3}
\end{figure}
\noindent
We now combine the events from $gg\to\gamma\gamma$ and $qX\to\gamma\gamma$ processes after multiplying their cross-sections with the respective $K$-factors [Eqs.~\eqref{eq:kgg}~\&~\eqref{eq:kpp}, respectively]. While combining we keep the luminosity same as the $qX$ processes, $346$ fb$^{-1}$. We then fit the combined events with the ATLAS fitting function [Eq.~\eqref{eq:fitform}].The dip due to the gluon fusion is clearly visible in the top panel of Fig.~\ref{fig:thres3} where we show the difference between the combined events and the ATLAS fit. As earlier, we focus in $250 \leq M_{\gamma\gamma} \leq 510$ GeV and fit the difference with a function that is a linear combination of $g_0(x)$ [Eq.~\eqref{eq:gg_fitform}] and a quadratic polynomial, \footnote{Increasing the degree of this polynomial does not lead to any better fit.} 
\begin{eqnarray}
c_0(x)&=& \mathcal{G}\Big[-\exp\Big\lbrace-\frac{(x-x_0)^2}{2\sigma_{g}^2}\Big\rbrace 
+ \mathcal R \Big\lbrace \exp\left(-\frac {x}{ \sigma} \right)\nonumber\\
&&-\left(\frac{\sigma}{x}\right)^4\Big\rbrace\Big]
+\Big(\mathcal{Q}_0 + \mathcal{Q}_1 x +\mathcal{Q}_2 x^2\Big).
\label{eq:combined_fitform}
\end{eqnarray}
The quadratic polynomial is added to account for the fluctuations
of the $qX\rightarrow\gamma\gamma$ process seen in Fig.~\ref{fig:thres2}. Note that in this step, only $\mathcal Q_i$'s and $\mathcal G$ are allowed to vary while other parameters ({\it i.e.} $\mathcal R$, $\sigma$, $\sigma_g$ and $x_0$) are held fixed to their respective values obtained earlier. 
In the bottom panel of Fig.~\ref{fig:thres3} we show the fit thus obtained. The parameter values are $\mathcal G=624.35$, $\mathcal Q_0=-1.6684\times10^{3}$, $\mathcal Q_1=1.1332\times{10}^{5}$ and $\mathcal Q_2=-1.8397\times10^{6}$. This four parameter fit has $\chi^2/{\rm d.o.f.} = 1.09$, indicating that it is a good fit. The one sigma error in $\mathcal G$ is about $114$, which indicates that the dip is
detected at the significance of roughly 5$\sigma$. 
This demonstrates that within $346~{\rm fb}^{-1}$ of LHC luminosity such a statistical fitting can isolate the gluon fusion threshold contribution with a high degree of accuracy.
A reasonable three to four sigma detection may be possible even with much lower luminosity. Hence
it would be worthwhile to apply our analysis to data that would be
available in near future.

The above procedure of fitting the difference effectively amounts to fitting the $pp\to\gamma\gamma$ combined events with the following function
\begin{eqnarray}
f^\prime_0(x) = \left\{\begin{array}{ll}
f_0(x) + c_0(x) & \mbox{ if } 250\leq M_{\gamma\gamma}\leq 510 \mbox{ GeV}\\&\\
f_0(x) & \mbox{ otherwise}
\end{array}\right.
\end{eqnarray}
instead of fitting them with only $f_0(x)$. Note that this increases the number of parameters in the fit just by four ($\mathcal Q_i$'s and $\mathcal G$, all other parameters inside $g_0(x)$ are already determined before). In the full range, $200\leq M_{\gamma\gamma} \leq 1000$ GeV, the ATLAS fit (two parameters) has $\chi^2_{global}=64$ while the fit with $f^\prime_0(x)$ (six parameters) has $\chi^2_{global}=38$.
Hence it is clear that our fitting function provides a much improved fit to the data.

In this paper, we have focused on detecting the signal of the dip. However, as
pointed out in the literature \cite{Li:2011ye,Campbell:2016yrh,Chway:2015lzg,Kawabata:2016aya}, it may provide a useful measurement of the top quark mass. From  Fig.~\ref{fig:thres3} we see that it may be possible to extract
useful information about the top mass from this analysis. The main
challenge would be to minimize the error induced by the fluctuations due
to the $qX\rightarrow \gamma\gamma$ process. We postpone a detailed analysis of such an
extraction to future research.

\section{Possible Improvements}
\noindent
So far, apart from applying the global cuts shown in Eq.~\eqref{eq:ATLAScuts}, we have not used any other technique to improve the $\sigma_{gg}/\sigma_{qX}$ ratio. It is actually quite difficult to isolate the $gg$ channel from the $qX$ significantly using simple kinematic cuts -- most of the $gg$ distributions (of standard variables $p_{\rm T}, \eta$ of photons etc.) are very similar to those of $qX$. The small cross-section of the $gg$ channel complicates the situation further. In such a situation, sophisticated numerical techniques (like multivariate analysis with machine learning) could possibly help, but are beyond the scope of this paper.  Here we sketch some such possible directions with some basic estimation.

\begin{table}[b]
\centering
\caption{Comparison of cross-sections in the $gg$ and the $q\bar q$ channels after application of various selection criteria.  Criterion $\mathcal C_1$ denotes the kinematic selection cuts [Eq.~\eqref{eq:ATLAScuts}], $\mathcal C_2$ stands for selection of events that pass through $\mathcal C_1$ and have a gluon jet as the leading $(p_T)$ jet.}
\label{tab:table1}
\begin{tcolorbox}[tab2,tabularx={c||Y|Y|Y}]
{\rm\footnotesize Selection Criteria}&   {$\sigma_{gg}(fb)$} & {$\sigma_{qX}(fb)$} & $\sigma_{gg}/\sigma_{qX}$\\\hline\hline
$\mathcal C_1$& $130$ & $1718$& $0.07$\\\hline
$\mathcal C_2$& $41$ & $350$ & $0.12$
\end{tcolorbox}
\end{table}

\begin{figure}[t]
\begin{center}
\scalebox{0.48}{\includegraphics*[angle=0,width=1.0\textwidth,clip]{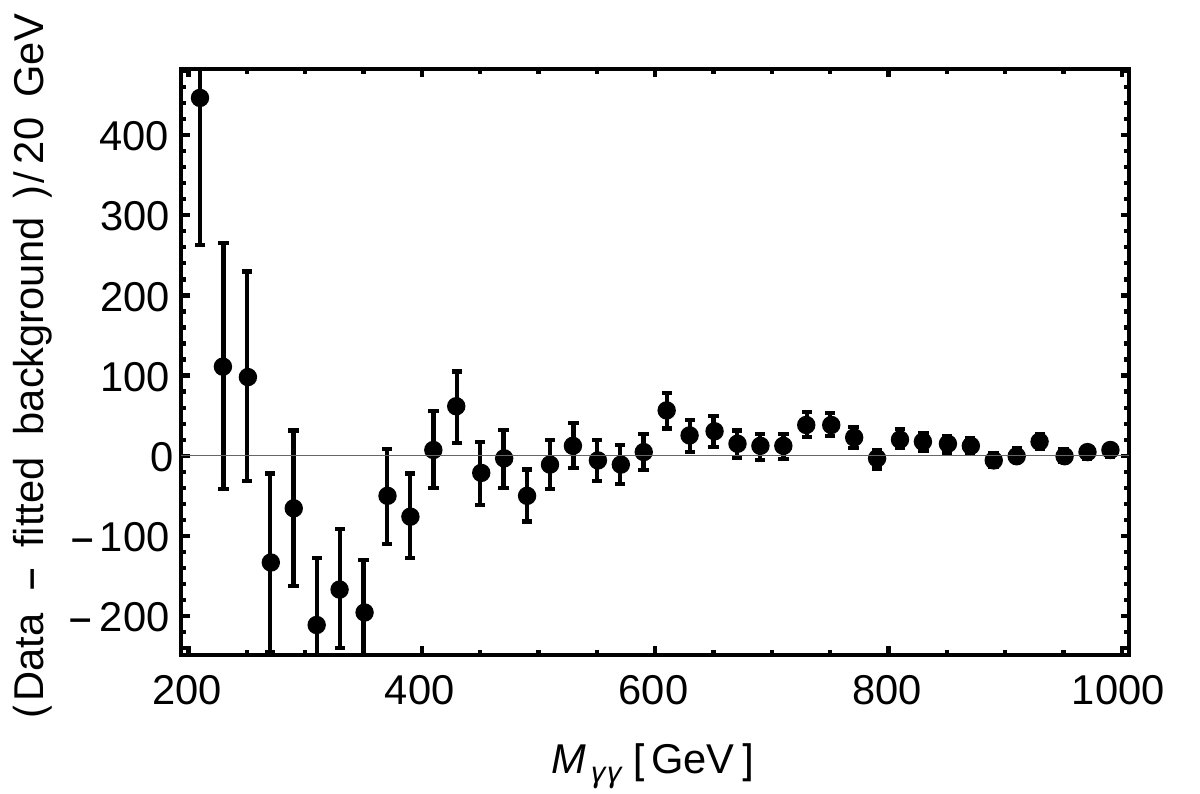}}
\end{center}
\caption{The difference between the binned events with selection criteria $\mathcal{C}_2$ and their smooth fit by the ATLAS fitting function [Eq.~\eqref{eq:fitform}] for the combined $gg\rightarrow\gamma\gamma + qX\rightarrow\gamma\gamma$ process. 
}
\label{fig:thres4}
\end{figure}

In the literature there are several studies showing that it is possible to statistically discriminate between a gluon-jet and a quark jet \cite{CMS-DP-2016-070, CMS-PAS-JME-13-002, Gallicchio:2011xq, Gallicchio:2012ez,Altheimer:2012mn,Cheng:2017rdo}. Now, since, a good fraction of total $qX$ events will have a  leading quark jet (because of the presence of $qg\to\gamma\gamma q$), if we demand the leading $p_{\rm T}$ jet not to be a quark jet, it should improve the $\sigma_{gg}/\sigma_{qX}$ ratio. The second row of Table \ref{tab:table1} shows that such a condition does indeed improve the ratio, even though the cross-sections reduce significantly. At best, we only find a marginal improvement
with this condition. In Fig.~\ref{fig:thres4}, we show the difference obtained from the same combined events but with this new condition of jet flavor on the leading jet.

While obtaining the above estimate we have ignored the efficiency of gluon-jet tagging. In other words,  we have been optimistic and have taken the gluon-jet-tagging efficiency, $\epsilon =1$. In practice, not all quark/gluon-jets can be identified and also a fraction of quark jets will be misidentified as gluon jets, making $\epsilon <1$.  This will increase the luminosity requirement. For simplicity, if we assume $\epsilon$ is same for the $qX$ and $gg$ channels and set a modest value, {\it i.e.}, $\epsilon = 0.4$ (current CMS estimations put the gluon tagging efficiency close to 40\% while quark identification efficiency to about 70\% \cite{CMS-DP-2016-070, CMS-PAS-JME-13-002}), the luminosity required to observe the dip with the same significance will be quite high, $346/\epsilon = 865$ fb$^{-1}$. However, one has to keep in mind that the obtained luminosity requirement is an estimate; advanced analysis technique like multivariate analysis can reduce the luminosity requirement significantly. Also, the recent advancements in jet-substructure techniques indicate 
to the possibility of significant improvement of quark/gluon tagging efficiency.

Before we proceed further, we would like to mention that we have also explored the possibility of varying the $M_{\gamma\gamma}$ dependent cut in Eq.~\eqref{eq:ATLAScuts} to improve the significance. We have tried loosening the cuts on the transverse energies of the photons even further, such as,
\begin{eqnarray}
E^{\gamma_1}_{\rm T} \geq E^{\gamma_2}_{\rm T}  > 40 \hspace{1mm}\text{GeV},\quad 
\end{eqnarray}
\begin{eqnarray}
E^{\gamma_1}_{\rm T} > 40 \hspace{1mm}\text{GeV},\quad
E^{\gamma_2}_{\rm T} > 25 \hspace{1mm}\text{GeV}\quad 
\end{eqnarray}
and
\begin{eqnarray}
E^{\gamma_1}_{\rm T} > 25 \hspace{1mm}\text{GeV},\quad
E^{\gamma_2}_{\rm T} > 22 \hspace{1mm}\text{GeV}\quad\label{eq:Referee_Cut}
\end{eqnarray}
with $M_{\gamma\gamma}\geq200$ GeV. Out of these, the one in Eq.~\eqref{eq:Referee_Cut} gives the best results. For this cut we again generate $8000$ K and $600$ K events for the $qX$ and $gg$ channels respectively with $E_{\rm T}^\gamma > 20$ GeV and $|\eta^\gamma|<2.5$. However, with such loose cut on $E_{\rm T}^\gamma$, the $qX$ cross-section becomes huge and, as a result, even with such large number of events we can only probe a lower luminosity, about 50 fb$^{-1}$. On these events we apply the cut defined in Eq.~\eqref{eq:Referee_Cut} and follow the same steps described in Sections \eqref{eq:Sec1}, \eqref{eq:Sec2} and \eqref{eq:Sec3}. We get the cross-sections of $gg\rightarrow\gamma\gamma$ and $qX\rightarrow\gamma\gamma$ as 201 fb and 2690 fb respectively after multiplying with the respective $K$-factors (which are updated accordingly). For 50 fb$^{-1}$ of integrated luminosity, we obtain $\mathcal{G}=117 \pm 62$, {\it i.e.}, the dip is detected at a significance of roughly $2\sigma$. Our estimation indicates that to get the threshold effect at a significance of $4\sigma$ we need roughly the four times the number of events generated. However, performing such computation within a reasonable time limit is beyond our existing capabilities. Hence we refrain from analyzing this direction further.

\section{Other sources of threshold effect}
\noindent
Finally, before we conclude, we note that there are other sources
of top threshold effect in the photon pair production channel. 
Just like the $pp\to q\bar q\gamma\gamma$ process (that appears as a NNLO correction to the $pp\to\gamma\gamma$ process), one can consider $pp\to t \bar t\gamma\gamma$. This opens up when the center-of-mass energy in the parton frame crosses two times the mass of the top quark. However, the cross-section of this process is tiny compared to the processes considered here (about $5$ fb which further reduces to about $0.7$ fb once we set $M_{\gamma\gamma}\geq 200$ GeV) and it induces no noticeable feature in the $M_{\gamma\gamma}$ distribution. Then, just as the gluon splitting creates $c\bar c$ and $b\bar b$ pairs that contribute in the 
sea quark density of a proton, once the top threshold is crossed, one could also imagine $t\bar t$ pairs appearing in the sea ({\it i.e.} a `$t$-PDF') (see, {\it e.g.}, \cite{Han:2014nja}). Hence, additional threshold effects would arise due to the processes 
$t\bar t\rightarrow \gamma\gamma$ 
and $tg\rightarrow t\gamma\gamma$.
If, na\"{i}vely, one assumes that the `$t$-PDF' at a scale $Q$ is roughly given by the $b$-PDF at the scale $Qm_b/m_t$, the contributions of these processes  turn out to be much smaller than the effect considered here. It is not easy to quantify this argument, as the behavior of this density function near the top threshold won't be captured properly due to sizable top mass effects. 
However, since the top quark is much heavier than the $b$-quark, it is reasonable to 
assume that these processes are unlikely to produce any observable features with the present luminosity. 
Qualitatively, for the $tg\rightarrow t\gamma\gamma$ process, this is supported by the small cross-section of the $pp\to t\bar t\gamma\gamma$ process mentioned before.

\section{Summary and Conclusions}
\noindent
In this paper we have investigated the top quark threshold effect in the two photon channel at the LHC. 
This effect arises in the loop mediated $gg\to\gamma\gamma$ subprocess due to the destructive interference
between top loop diagrams containing on-shell top quarks and other light quark loop diagrams. It
appears as a dip in the invariant mass distribution of the photon pair near two times the mass of the
top quark.

Within the SM, the gluon fusion 
process is overshadowed by the $qX\to\gamma\gamma$ process that has a larger cross-section.
However, here, we have argued that a statistical fit can isolate the dip in the gluon channel reasonably well.
Though it is beyond the scope of 
this paper, it might be possible to make the threshold effect even more prominent with sophisticated techniques like 
multivariate analysis etc.

 It will be very interesting to observe this SM effect more accurately at the LHC.  However, there are other motivations to look into this in detail. 
It can provide us yet another way to probe the top-mass experimentally. Not only this, threshold effects can also tell us about some beyond the SM fermions indirectly.  
Since this effect arises from the interference effects, any heavy fermion that can run in the loop
of the $gg\to\gamma\gamma$ process would lead to such a threshold effect. Hence, observation (or non-observation) of any such effect
in the gamma gamma spectrum could let us infer about new heavy colored fermions carrying non-zero electromagnetic charge in a model independent manner. 

\section{Acknowledgments}
\noindent
PS and RKV thank the Ministry of Human Resource Development (MHRD), Government of India for their Ph.D. fellowships. We thank an anonymous referee for useful comments which have helped to
improve the paper.

\bibliographystyle{JHEP}
\bibliography{Reference}

\end{document}